\documentclass{article}[12pt]
\usepackage{graphicx} 
\usepackage{indentfirst}
\usepackage{geometry}
\usepackage{setspace}
\usepackage{url}
\usepackage{hyperref}
\usepackage{color}

\title{Reply to: Deep reinforced learning heuristic tested on spin-glass ground states: The larger picture}
\author{\small Changjun Fan, Mutian Shen, Zohar Nussinov, Zhong Liu, Yizhou Sun, Yang-Yu Liu}

\begin{document}
\maketitle
We wish to thank Stefan Boettcher for prompting us to further check and highlight the accuracy and scaling of our results~\cite{fan_searching_2023}. Here we provide a comprehensive response to the Comment written by him~\cite{boettcher_deep_2023}. We argue that the Comment did not account for the fairness of the comparison between different methods in searching for the spin-glass ground states. We demonstrate that, with a reasonably larger number of initial spin configurations, our results agree with the asymptotic scaling form assumed by finite-size corrections. 

\section*{3D Edwards–Anderson (EA) model}
\begin{figure}[htb]
    \centering
    \includegraphics[width=\textwidth]{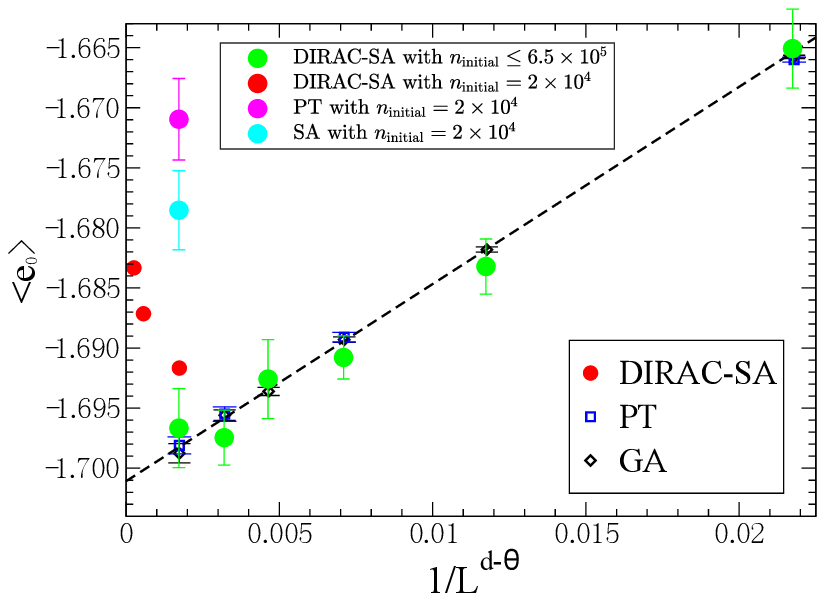}
    \caption{
    With a reasonably large $n_{\rm initial}$, our DIRAC-SA results agree well with larger picture suggested by FSC. 
    Our results are plotted on top of Fig. 1 from \cite{boettcher_deep_2023}. FSC assumes that the average ground state energy per spin of a given $d$-dimensional EA system of size $N=L^d$ has the form $\left<e_0\right>_{N} = \left<e_0\right>_{N=\infty} + A x + \cdots$, where $x=1/L^{d-\theta}$ and $d-\theta\approx 2.76$. Ignoring the higher order terms, this form is shown as the dashed line here. 
    The red, magenta and cyan points are $\left<e_0\right>_{N}$ for $N=10^3$ computed by DIRAC-SA, PT, and SA, respectively, all with $n_{\rm initial}=2.0\times 10^4$. 
    The green points represent $\left<e_0\right>_{N}$ for $N=4^3,5^3,6^3,7^3,8^3,10^3$, with $n=850,900,820,120,221,50$ instances respectively, computed by DIRAC-SA with $n_{\rm initial}\leq 6.5\times 10^5$. 
    }
    \label{fig:fig1}
\end{figure}

In Fig.5 of our paper~\cite{fan_searching_2023}, we plotted the disorder-averaged energy per spin (denote as $e_0$) as a function of the number of initial spin configurations (denoted as $n_{\rm initial}$) for different methods to benchmark those methods on large 3D EA Ising spin glass instances with Gaussian disorder. 
The Comment pointed out that DIRAC-SA (a variant of our DIRAC method) did not reach the ground states for those systems, as indicated by the large deviation of the three red points from the asymptotic scaling form assumed by finite-size corrections (FSC), see Fig.1 of the Comment and this response letter. However, as we explicitly mentioned in the caption of Fig.5 in our paper ~\cite{fan_searching_2023}, we only ran all the tested algorithms up to a small $n_{\rm initial}=2.0\times 10^4$, which is much smaller than the number required to reach the ground state, as reported in the literature. For instance, Ref.~\cite{ComparingMethods} reported that, to reach the ground state for $3D$ $L=10$ systems, the parallel tempering (PT) method requires $n_{\rm initial}=3.2\times 10^7$, which is $1,600$ times larger than the number of initial spin configurations we used. Such a big difference in terms of $n_{\rm initial}$ is certainly not inconsequential. We did not expect any of the methods to reach the ground state with $n_{\rm{initial}} = 2.0 \times 10^4$ for large 3D EA instances with Gaussian disorder. 
Indeed, for $3D$ $L=10$ systems, with $n_{\rm initial}=2.0\times 10^4$, PT and simulated annealing (SA) did not reach the expected ground state either (see the magenta and cyan points in Fig.\ref{fig:fig1} of this response). In fact, with the same $n_{\rm initial}$, results of these two methods are even farther away from the FSC line than DIRAC-SA for $3D$ $L=10$ systems (see the third red point in Fig.\ref{fig:fig1} of this response). 
Without specifying the number of initial spin configurations, we think it is unfair and meaningless to compare different methods in searching for the ground states of large spin-glass instances.  

In our paper~\cite{fan_searching_2023} we did not try a larger $n_{\rm{initial}}$ for two reasons. First, we had already demonstrated the ability of DIRAC to reach the exact ground states for small systems (which can be confirmed by the branch-and-bound-based solver Gurobi), as shown in Fig.4 of our paper~\cite{fan_searching_2023}. Second, we did not find it necessary to invest extensive computational resources in an ``arms race" fashion of computing the ``ground states" of these large systems for which exact solvers cannot confirm the results. Also, to achieve the (true) ground states the required $n_{\rm{initial}}$ may be exponential in the system size. There is no exception for DIRAC or any other heuristic methods.
Our paper aimed to demonstrate the effectiveness and efficiency of DIRAC over other methods at the same $n_{\rm{initial}}$, rather than to confirm the asymptotic scaling form assumed by FSC. We appreciate the ``larger picture" mentioned in the Comment. But it was beyond the scope of our paper. 

Since the Comment questioned the ability of our method to reach the ground state for large systems, we think it is necessary to perform heavier computations with a larger $n_{\rm initial}$ to directly address the Comment. For $3D$ $L=10$ systems with $n=50$ instances, we found that, with $n_{\rm initial}=6.5\times 10^5$, about $2\%$ as that needed for PT, the average energy per spin computed by DIRAC-SA could indeed reach the asymptotic scaling form assumed by FSC (see the leftmost green point in Fig.\ref{fig:fig1}). We also plotted $e_0$ computed by DIRAC-SA for $3D$ $L=4,5,6,7,8$, with $n=850,900,820,120,221$ instances respectively, in the same figure. We found that they agree well with the FSC line. These results clearly demonstrate that the importance of using a large $n_{\rm{initial}}$ to achieve results consistent with the prediction of FSC. We are grateful that the Comment helped us clarify this point. As mentioned above, confirming the asymptotic scaling form assumed by FSC was not the original goal of our paper.  

\section*{Sherrington–Kirkpatrick (SK) model}
Fig.2 of the Comment acknowledged that our results for the SK model are consistent with the asymptotic scaling form assumed by FSC, although in the figure we could still see a deviation from the FSC line for SK model of $N=64$. We believe this deviation is simply due to the small number of instances ($n=50$) used in our calculation. We notice that with $n=50$ instances the results offered by the extremal optimization (EO) heuristic also deviate from the FSC line, especially for $N=125$. We argue that DIRAC needs more instances to reach the FSC line, just like the EO case. After all, only the average over many different instances may be expected to behave as a smooth function of $N$~\cite{1pal_ground_1996}.

The Comment also pointed out that the system sizes we considered are relatively small. We emphasize that, as a reinforcement-learning framework based on graph neural network, DIRAC was not specifically designed for SK models with a complete graph topology. 
We believe that, to compute ground states for larger SK instances, DIRAC would have to be modified to explicitly consider the complete graph topology. 
However, this was beyond the scope of our paper. 

\section*{Competitive methods}
It is a pity that in our paper we did not explicitly cite any papers on the genetic algorithm~\cite{1pal_ground_1996,1pal_ground_1996-1} (GA) or extremal optimization (EO) heuristic~\cite{boettcher_extremal_2005,boettcher_optimization_2001,middleton_improved_2004}. We did cite a book\cite{hartmann2004new} on the use of those heuristic methods for computing the spin-glass ground state though, as also pointed out by the Comment.  
In our paper, we did not compare the performance of DIRAC with that of GA and EO either. 
This is mainly because PT and GA were commonly used to compute the ground state of the EA Ising spin glass model with Gaussian disorder \cite{roma2009ground,ComparingMethods}, and Ref.\cite{roma2009ground} reported that a simple PT algorithm performs as well as GA found in the literature. Hence, we chose PT as a competitive method of DIRAC.  
We did consider two classical heuristic methods: SA and Greedy algorithm. 
Overall, we think comparing DIRAC with those methods is sufficient to demonstrate its superiority.

\vspace{0.1in}
\noindent{\bf Acknowledgements:}  We wish to thank Stefan Boettcher for discussions and correspondence.


\bibliographystyle{unsrt}
\bibliography{ref}

\begin{thebibliography}{10}

\bibitem{fan_searching_2023}
Changjun Fan, Mutian Shen, Zohar Nussinov, Zhong Liu, Yizhou Sun, and Yang-Yu
  Liu.
\newblock Searching for spin glass ground states through deep reinforcement
  learning.
\newblock {\em Nature Communications}, 14(1):725, February 2023.
\newblock Number: 1 Publisher: Nature Publishing Group.

\bibitem{boettcher_deep_2023}
Stefan Boettcher.
\newblock Deep reinforced learning heuristic tested on spin-glass ground
  states: {The} larger picture, February 2023.
\newblock arXiv:2302.10848 [cond-mat].

\bibitem{ComparingMethods}
Wenlong Wang, Jonathan Machta, and Helmut~G. Katzgraber.
\newblock Comparing monte carlo methods for finding ground states of ising spin
  glasses: Population annealing, simulated annealing, and parallel tempering.
\newblock {\em Physical Review E}, 92:013303, Jul 2015.

\bibitem{1pal_ground_1996}
Károly~F. Pál.
\newblock The ground state of the cubic spin glass with short-range
  interactions of {Gaussian} distribution.
\newblock {\em Physica A: Statistical Mechanics and its Applications},
  233(1):60--66, November 1996.

\bibitem{1pal_ground_1996-1}
Károly~F. Pál.
\newblock The ground state energy of the {Edwards}-{Anderson} {Ising} spin
  glass with a hybrid genetic algorithm.
\newblock {\em Physica A: Statistical Mechanics and its Applications},
  223(3):283--292, January 1996.

\bibitem{boettcher_extremal_2005}
S.~Boettcher.
\newblock Extremal optimization for {Sherrington}-{Kirkpatrick} spin glasses.
\newblock {\em The European Physical Journal B: Condensed Matter and Complex
  Systems}, 46(4):501--505, 2005.
\newblock Publisher: Springer \& EDP Sciences.

\bibitem{boettcher_optimization_2001}
Stefan Boettcher and Allon~G. Percus.
\newblock Optimization with {Extremal} {Dynamics}.
\newblock {\em Physical Review Letters}, 86(23):5211--5214, June 2001.
\newblock Publisher: American Physical Society.

\bibitem{middleton_improved_2004}
A.~Alan Middleton.
\newblock Improved extremal optimization for the {Ising} spin glass.
\newblock {\em Physical Review E}, 69(5):055701, May 2004.
\newblock Publisher: American Physical Society.

\bibitem{hartmann2004new}
Alexander~K Hartmann and Heiko Rieger.
\newblock New optimization algorithms in physics.
\newblock 2004.

\bibitem{roma2009ground}
F~Rom{\'a}, S~Risau-Gusman, Antonio~Jose Ramirez-Pastor, F~Nieto, and Eugenio~E
  Vogel.
\newblock The ground state energy of the edwards--anderson spin glass model
  with a parallel tempering monte carlo algorithm.
\newblock {\em Physica A: Statistical Mechanics and its Applications},
  388(14):2821--2838, 2009.

\end{thebibliography}
\end{document}